\documentclass[10pt]{article}
\usepackage{fullpage}
\usepackage{amsmath}
\usepackage{amssymb}
\usepackage[dvips]{epsfig}

\usepackage{amssymb}
\usepackage{amsmath}
\usepackage{epsf}
\usepackage{graphicx}
\usepackage[dvips]{epsfig}
\usepackage{bm}
\usepackage{color}

\def\##1{{\bf #1}}
\def\=#1{\underline{\underline{#1}}}

\def\+#1{\underline{\bf #1}}
\def\*#1{\underline{\underline{\bf #1}}}

\def\r#1{(\ref{#1})}
\def\l#1{\label{#1}}
\def\c#1{\cite{#1}}

\def\le{\left(}
\def\ri{\right)}
\def\les{\left[}
\def\ris{\right]}
\def\lec{\left\{}
\def\ric{\right\}}

\def\.{\mbox{ \tiny{$^\bullet$} }}

\def\epso{\epsilon_{\scriptscriptstyle 0}}

\def\muo{\mu_{\scriptscriptstyle 0}}

\def\ko{k_{\scriptscriptstyle 0}}

\def\co{c_{\scriptscriptstyle 0}}

\def\eps{\epsilon}

\begin{document}

\begin{center}

{\bf {\LARGE Negative and positive refraction are not Lorentz
covariant }}

\vspace{10mm} \large

 Tom G. Mackay\footnote{E--mail: T.Mackay@ed.ac.uk.}\\
{\em School of Mathematics and
   Maxwell Institute for Mathematical Sciences\\
University of Edinburgh, Edinburgh EH9 3JZ, UK}\\
and\\
 {\em NanoMM~---~Nanoengineered Metamaterials Group\\ Department of Engineering Science and Mechanics\\
Pennsylvania State University, University Park, PA 16802--6812,
USA}\\
 \vspace{3mm}
 Akhlesh  Lakhtakia\footnote{E--mail: akhlesh@psu.edu}\\
 {\em NanoMM~---~Nanoengineered Metamaterials Group\\ Department of Engineering Science and Mechanics\\
Pennsylvania State University, University Park, PA 16802--6812, USA}

\end{center}

\vspace{4mm}

\normalsize

\begin{abstract}

The refraction of linearly polarized plane waves into a half--space
occupied by a  material moving at constant velocity was studied by
directly implementing the Lorentz transformations of electric and
magnetic fields. From the perspective of a co--moving observer, the
moving material was a spatially local, pseudochiral omega material.
Numerical studies revealed that whether or not negative refraction
occurrs in the moving material depends upon the  speed of movement
as well as the angle of incidence and the polarization state of the
incident plane wave. Furthermore, the phenomenons of negative phase
velocity and counterposition in the moving material were similarly
found not to be Lorentz covariant; both phenomenons were also
found to be sensitive to the angle
of incidence and the polarization state of the incident plane wave.

\end{abstract}


% counterposition, negative phase velocity,
%Lorentz transformations, Minkowski constitutive relations, Poynting
%vector

\section{Introduction}

The behaviour of plane waves at a planar interface between two
disparate homogeneous mediums  is a central topic in both
fundamental and applied electrodynamics. In particular, the
phenomenon of negative refraction \c{Laeu,EJP,SAR} has been the
subject of intense research efforts for the past ten years,
following experimental reports of this phenomenon in certain
metamaterials \c{Smith_PRL,Shelby}. Much of this effort has been
motivated by the development of novel metamaterials, but negative
refraction also arises in certain minerals \c{Pimenov} and in
biological structures \c{Stavenga}. In addition, the prospects of
negative refraction arising in relativistic scenarios~---~such as in
uniformly--moving materials \c{ML_Optik,ML_JPA_sub} or in strong
gravitational fields
\c{ML_JPA_GTR,MLS_NJP2,Sharif1,Komissarov}~---~is a matter of
 astrophysical and astronomical significance.

In the following we consider  negative refraction induced by uniform
motion. Earlier work relating to this topic relied on the Minkowski
constitutive relations to describe the moving medium in a
nonco--moving inertial reference frame
\c{Kong_PRB,MOTL_counterposition,Hiding}, per the standard textbook
approach \c{Chen}. However, this approach is only appropriate to
materials which are both spatially and temporally local \c{BC}. More
recently, studies based on the uniform motion of realistic materials
have been undertaken, using an approach in which the Lorentz
transformations of the electric and magnetic fields are directly
implemented \c{ML_Optik,ML_JPA_sub}. These studies revealed that the
phenomenons of negative phase velocity and
counterposition\footnote{See Sec.~\ref{NR_NPV_C} for definitions of
negative refraction, negative phase velocity and
counterposition.}~---~which are closely allied to negative
refraction and similarly associated with certain metamaterials and
relativistic scenarios~---~are not Lorentz covariant. Here we
address the hitherto outstanding question: is negative (or positive)
refraction Lorentz covariant? By means of a numerical analysis based
on a uniformly moving pseudochiral omega material, we demonstrate
that the answer to this question is `no'.

In the notation we adopt, 3--vectors are in boldface with the
$\hat{}$ symbol denoting a unit vector. Double underlining signifies
a 3$\times$3 dyadic (i.e., a second rank Cartesian tensor) and the
identity 3$\times$3 dyadic is written as $\=I = \hat{\#x}\,\hat{\#x}
+ \hat{\#y}\,\hat{\#y} + \hat{\#z}\,\hat{\#z}$. The operators Re and
Im deliver the real and imaginary parts of complex quantities; and
$i = \sqrt{-1}$. The permittivity and permeability of free space are
$\epso$ and $\muo$, respectively, with $\co = 1 / \sqrt{ \epso
\muo}$ being the speed of light in free space.

\section{Refraction into a moving pseudochiral omega material} \l{moving_hs}

\subsection{Panewave analysis} \l{pw_analysis}

Our attention is focussed on
 a spatially local, homogeneous material,  characterized by the
 frequency--domain constitutive
 relations
\begin{equation}
\left. \begin{array}{l}
 \#D' = \=\eps' \. \#E' + \=\xi' \. \#H'
\vspace{6pt}\\
\#B' = \=\zeta' \. \#E' + \=\mu' \. \#H'
\end{array}
\right\} \l{CRs}
\end{equation}
in the inertial reference frame $\Sigma'$. Herein, the 3$\times$3
constitutive dyadics
\begin{eqnarray}
&& \=\eps' = \epso \le \begin{array}{ccc} \eps'_x & 0 & 0 \\
0 & \eps'_y & 0 \\
0 & 0 & \eps'_z \end{array}  \ri, \quad
\=\xi' = \frac{1}{\co} \le \begin{array}{ccc} 0 & 0 & 0 \\
0 & 0 & 0 \\
0 & -i \xi' & 0 \end{array} \ri, \nonumber \\
&& \=\zeta' = \frac{1}{\co} \le \begin{array}{ccc} 0 & 0 & 0 \\
0 & 0 & i \xi' \\
0 & 0 & 0 \end{array}  \ri, \quad
\=\mu' = \muo \le \begin{array}{ccc} \mu'_x & 0 & 0 \\
0 & \mu'_y & 0 \\
0 & 0 & \mu'_z \end{array} \ri. \label{cons}
\end{eqnarray}
 This is a
bianisotropic, Lorentz--reciprocal \c{Krowne}  material, known as a
\emph{pseudochiral omega material}  \c{Engheta_MOTL}. Constitutive
relations of this form have been used to describe certain
negatively refracting metamaterials, assembled from layers of
split--ring resonators \c{Chen_PRE}.  Several different designs of
metamaterials are based on this general configuration
\c{Wegener_Nature,Wegener_OL,Ozbay}. As the pseudochiral omega
material is presumed to be dissipative, the constitutive parameters
$\eps'_{x,y,z}$, $\xi'$ and $\mu'_{x,y,z}$ are complex--valued
functions of the angular frequency $\omega'$.

Suppose that  the  pseudochiral omega material fills the half--space
$z
> 0$, while the half--space $z < 0$ is vacuous.
The  inertial reference frame $\Sigma'$ translates at constant
velocity $\#v=v\hat{\#v}$ with respect  to the inertial reference
frame
 $\Sigma$, in the plane of the interface $z=0$. In keeping with an
 earlier study \c{ML_JPA_sub}, we take $\hat{\#v} = \hat{\#x}$.
  The Lorentz transformations \c{Chen}
\begin{equation}
\left. \begin{array}{l} \#E = \displaystyle{\le \#E' \.\hat{\#v} \ri
\hat{\#v} +
 \gamma \,
\les \le \=I - \hat{\#v}\hat{\#v} \ri \. \#E' - \#v \times
\#B' \ris } \vspace{4pt}  \\
\#B = \displaystyle{\le \#B' \.\hat{\#v} \ri \hat{\#v} +
 \gamma \,
\les \le \=I - \hat{\#v}\hat{\#v} \ri \. \#B' + \frac{ \#v \times
\#E'}{\co^2} \ris  } \vspace{4pt} \\
\#H = \displaystyle{\le \#H' \.\hat{\#v} \ri \hat{\#v} +
 \gamma \,
\les \le \=I - \hat{\#v}\hat{\#v} \ri \. \#H' + \#v \times
\#D' \ris  } \vspace{4pt} \\
\#D = \displaystyle{\le \#D' \.\hat{\#v} \ri \hat{\#v} +
 \gamma \,
\les \le \=I - \hat{\#v}\hat{\#v} \ri \. \#D' - \frac{ \#v \times
\#H'}{\co^2} \ris }\end{array} \right\},  \l{Dp}
\end{equation}
with the real--valued scalars
\begin{equation}
\gamma = \frac{1}{\sqrt{1 - \beta^2}},\qquad \beta =
\frac{v}{\co}\,,
\end{equation}
relate the  electromagnetic field phasors in the frame $\Sigma$  to
those in the frame $\Sigma'$.

Now suppose that the vacuous half--space $z<0$ contains a line
source, which is stationary with respect to the frame $\Sigma$. The
source extends infinitely in directions parallel to the $y$ axis,
and it is located at a great distance from the interface $z=0$. Let
us consider one plane wave incident on the interface $z=0$, as a
representative of the angular spectrum of plane waves launched by
the source. With respect to the frame $\Sigma$, this plane wave is
described  by the electric and magnetic field phasors
\begin{equation}
\left.\begin{array}{l}
\#E_i  = \#e_{i}\, \exp \les i \le \#k_{i}\cdot\#r - \omega t \ri \ris   \\[5pt]
\#H_i  = \#h_{i} \, \exp \les i \le \#k_{i} \cdot\#r   - \omega t
\ri \ris
\end{array}\right\}, \qquad z \leq 0.
\l{pw_vac}
\end{equation}
 Herein, the wavevector
\begin{equation}
\#k_{i} = \kappa \, \hat{\#x} + \ko \,\cos \theta \, \hat{\#z} ,
\end{equation}
with the real--valued scalar
 \begin{equation} \kappa = \ko\sin\theta \in \left( -\ko,
\, \ko \right),
\end{equation}
 the free--space wavenumber  $ \ko = \omega \sqrt{\epso \muo}$ and
$\omega$ being the angular frequency with respect to $\Sigma$.

With
respect to the frame
  $\Sigma'$, the incident plane wave is represented by
\begin{equation}
\left.\begin{array}{l}
\#E'_i  = \#e'_{i}\, \exp \les i \le \#k'_{i}\cdot\#r' - \omega' t' \ri \ris   \\[5pt]
\#H'_i = \#h'_{i} \, \exp \les i \le \#k'_{i} \cdot\#r'   - \omega'
t' \ri \ris
\end{array}\right\},\qquad z \leq 0,
\l{pw_vac_p}
\end{equation}
wherein the  phasor amplitudes $\lec \#e'_i, \#h'_i \ric$ are
related to $\lec \#e_i, \#h_i \ric$
  via the Lorentz transformations \r{Dp}, while
\begin{equation}
\left.
\begin{array}{l}
 \#k_{i} = \displaystyle{ \gamma \le \#k'_{i} \. \hat{\#v} + \frac{\omega' v}{\co^2}
\ri
\hat{\#v} + \le \, \=I - \hat{\#v} \, \hat{\#v} \ri \. \#k'_{i}} \vspace{6pt}\\
\#r = \displaystyle{\les \, \=I  + \le \gamma - 1 \ri
\hat{\#v}\,\hat{\#v} \ris \.
\#r' + \gamma \, \#v t'}\vspace{6pt} \\
 \omega = \displaystyle{\gamma \le \omega' + \#k'_{i} \. \#v \ri} \vspace{6pt}\\
 t = \displaystyle{\gamma \le t' + \frac{\#v \. \#r'}{\co^2} \ri}
 \end{array} \right\}. \l{wT}
 \end{equation}

The incident plane wave gives rise to two refracted plane waves in
the half--space $z > 0$, and one reflected plane wave in the
half--space $z< 0$. In the frame $\Sigma'$, the refracted plane
waves are represented  by the electric and magnetic phasors
\begin{equation}
\left.\begin{array}{l}
\#E'_t  = \#e'_{tj}\, \exp \les  i \le \#k'_{tj}\cdot\#r' - \omega' t' \ri \ris  \\[5pt]
\#H'_t  = \#h'_{tj} \, \exp\les i \le  \#k'_{tj} \cdot\#r'   -
\omega' t' \ri \ris
\end{array}\right\}, \qquad z \geq 0,
\qquad (j=1,2) , \l{pw_mat}
\end{equation}
wherein the wavevectors
\begin{equation}
\#k'_{tj} =  \left(\#k'_i \.\hat{\#x}\right) \hat{\#x} + k'_{zj} \,
\hat{\#z}, \qquad (j=1,2)
\end{equation}
comply with  Snel's law \c{Chen}. The wavevector components
$k'_{zj}$, as well as the relationships between the phasor
amplitudes $\#e'_{tj}$ and $\#h'_{tj}$,
 are deduced by combining the
 constitutive relations
\r{CRs} with the source--free Maxwell curl postulates in $\Sigma'$
\c{Chen}. We find \c{ML_PRB}
 \begin{equation}
 \left.
 \begin{array}{l}
 k'_{z1} = \displaystyle{  \omega' \sqrt{\epso \muo} \sqrt{  \mu'_x \le \eps'_y - \frac{ \left(\#k'_i
  \.\hat{\#x}\right)^2}{\omega'^{\,2} \epso \muo  \mu'_z }
 \ri}} \vspace{6pt}\\
k'_{z2} = \displaystyle{ \omega' \sqrt{\epso \muo}
\sqrt{\frac{\eps'_x}{\eps'_z} \les  \le \eps'_z \mu'_y - \xi'^{\,2}
\ri - \frac{ \left(\#k'_i \.\hat{\#x}\right)^2}{\omega'^{\,2} \epso
\muo } \ris }}
\end{array}
\right\}. \label{k1k2}
\end{equation}
Notice that since $k'_{zj}$ are generally complex--valued, the
refracted plane waves are nonuniform.

The reflected plane wave is
represented by the  electric and magnetic phasors
\begin{equation}
\left.\begin{array}{l}
\#E'_r  = \#e'_{r}\, \exp \les i \le \#k'_{r}\cdot\#r' - \omega' t' \ri \ris   \\[5pt]
\#H'_r = \#h'_{r} \, \exp \les i \le \#k'_{r} \cdot\#r'   - \omega'
t' \ri \ris
\end{array}\right\},\qquad z \leq 0,
\l{pwr_vac_p}
\end{equation}
in the frame $\Sigma'$, with the  wavevector of the reflected plane wave being
\begin{equation}
\#k'_{r} =  \left(\#k'_i \.\hat{\#x}\right) \hat{\#x} - k'_{zr} \,
\hat{\#z}\,.
\end{equation}
The source-free Maxwell curl postulates in $\Sigma'$ yield an
expression for $k'_{zr}$ and relationships between the phasor
amplitudes $\#e'_{r}$ and $\#h'_{r}$.

By invoking the standard
 boundary conditions  across the plane $z=0$, i.e., \c{Chen}
\begin{equation} \l{bcs}
\left.\begin{array}{ll} (\#e'_i+ \#e'_r)\cdot\hat{\#x}=
\#e'_{tj}\cdot\hat{\#x}\, &
(\#e'_i+ \#e'_r)\cdot\hat{\#y}= \#e'_{tj}\cdot\hat{\#y}\\[5pt]
(\#h'_i+ \#h'_r)\cdot\hat{\#x}= \#h'_{tj}\cdot\hat{\#x}\, & (\#h'_i+
\#h'_r)\cdot\hat{\#y}= \#h'_{tj}\cdot\hat{\#y}
\end{array}\right\},\qquad z =0, \qquad (j=1,2),
\end{equation}
 the  phasor amplitudes $\lec \#e'_r, \#h'_r \ric$ and $\lec
\#e'_{tj}, \#h'_{tj} \ric$ can be found. The reflected and the
refracted plane waves in the frame $\Sigma$
 may then be  deduced by
applying the Lorentz transformations \r{Dp} and \r{wT}.

We represent
the refracted plane wave  in the frame $\Sigma$ by the electric and
magnetic field phasors
\begin{equation}
\left.\begin{array}{l}
\#E_t  = \#e_{tj}\, \exp \les  i \le \#k_{tj}\cdot\#r - \omega t \ri \ris  \\[5pt]
\#H_t  = \#h_{tj} \, \exp\les i \le  \#k_{tj} \cdot\#r   - \omega t
\ri \ris
\end{array}\right\}, \qquad z \geq 0, \qquad (j=1,2),
\l{pw_mat2}
\end{equation}
wherein the wavevectors
\begin{equation}
\#k_{tj} =   \kappa \hat{\#x} + k_{zj} \, \hat{\#z}, \qquad (j=1,2),
\end{equation}
while the corresponding reflected plane wave is represented by
\begin{equation}
\left.\begin{array}{l}
\#E_r  = \#e_{r}\, \exp \les  i \le \#k_{r}\cdot\#r - \omega t \ri \ris  \\[5pt]
\#H_r  = \#h_{r} \, \exp\les i \le  \#k_{r} \cdot\#r   - \omega t
\ri \ris
\end{array}\right\},
\l{pw_ref}
\end{equation}
with the wavevector
\begin{equation}
\#k_{r} =   \kappa \, \hat{\#x} - \ko \,\cos \theta \, \hat{\#z}.
\end{equation}

\subsection{Negative refraction, negative phase velocity and
counterposition} \l{NR_NPV_C}

We highlight three  phenomenons which the uniformly moving pseudochiral omega
medium can support: namely, negative refraction, negative phase
velocity and counterposition. All three phenomenons are closely
associated with certain metamaterials \c{ML_PRB}. We consider these
phenomenons referred to the frame $\#\Sigma$.

First, in the scenario under consideration here, negative refraction
occurs when the real part of the wavevector $\#k_{tj}$ casts a
negative projection onto the positive axis $z$; i.e., when ${\rm Re}
\, k_{zj} < 0$ \c{EJP}.
 Second, the phase velocity of the refracted
wave is classified as negative when the real part of the wavevector
$\#k_{tj}$ casts a negative projection onto the corresponding
cycle--averaged Poynting vector $\#P_j$  \c{Laeu}. For our purposes
here, it suffices to consider $\#P_j$ in the limit $| \#r | \to 0$,
as provided by
\begin{eqnarray}
 \left. \#P_{j} \, \right|_{| \#r | \to 0}  &=&    \frac{1}{2} \les \le \mbox{Re} \,
 \#e_{tj} \ri
\times \le \mbox{Re} \, \#h_{tj} \ri  + \le \mbox{Im} \, \#e_{tj}
\ri \times \le \mbox{Im} \, \#h_{tj} \ri \ris, \qquad (j=1,2).
\end{eqnarray}
The signs of the square roots in eqs.~\r{k1k2} are selected in order
to ensure that $\#P_j$ casts a positive projection onto the positive
$z$ axis. Thereby, energy flow directed  into the $z > 0 $
half--space is assured.
 Third,
counterposition occurs when the real part of the wavevector
$\#k_{tj}$ and the corresponding cycle--averaged Poynting vector
$\#P_j$  are oriented on opposite sides to the normal to the
interface $z=0$ \c{ZFM,Optik_counterposition,note1}. Because here
$\#P_j \. \hat{\#z}
> 0$, the conditions for counterposition are
\begin{equation} \l{counter_cond}
\left.
\begin{array}{lcr}
\le \mbox{Re} \,\#k_{tj} \.\hat{\#z} \ri  \le \, \#P_j \. \hat{\#x}
\,
\ri <0 & & \mbox{for} \: \kappa > 0 \vspace{6pt} \\
\le \mbox{Re} \,\#k_{tj} \.\hat{\#z} \ri  \le \, \#P_j \. \hat{\#x}
\, \ri > 0 & & \mbox{for} \: \kappa < 0
\end{array}
\right\}, \qquad (j=1,2).
\end{equation}

For uniform planewave propagation in an isotropic dielectric
material, negative refraction and negative phase velocity are
effectively synonymous \c{Laeu,EJP}. However, this is not the case
for  more complex materials \c{Belov_MOTL}. In particular, for the
pseudochiral omega material considered here, it has been established
that negative refraction, negative phase velocity and
counterposition are independent phenomenons \c{ML_PRB}.

\subsection{Numerical studies} \l{num_ex}

We now explore the phenomenons of negative refraction, negative
phase velocity and counterposition for  a specific numerical
example, chosen to allow direct comparison with an earlier study
\c{ML_PRB}. Let the pseudochiral omega medium occupying the
half--space $z > 0$ be characterized by the constitutive parameters:
$\eps'_x = 0.1 + 0.03i,$ $\eps'_y = 0.14 + 0.02i,$  $ \eps'_z = 0.13
+ 0.07i;$ $ \mu'_x = -0.29 + 0.09i,$ $\mu'_y = -0.18 + 0.03i,$  $
\mu'_z = -0.17 + 0.6i;$ and $ \xi' = 0.11 + 0.05i$.

 In the following
we consider two polarization states for the incident plane wave. As
described in the Appendix, the incident $s$--polarization state, as
characterized by
\begin{equation}
\#e_i = a_s \, \#s \equiv a_s \, \hat{\#y}, \l{s_polar}
\end{equation}
gives rise to the refracted plane wave with wavevector $\#k_{t1}$,
whereas the incident $p$--polarization state, as characterized by
\begin{equation}
\#e_i = a_p \,  \#p_+ \equiv a_p \le -   \cos \theta \, \hat{\#x} +
\sin \theta \, \hat{\#z} \ri ,
\end{equation}
gives rise to the refracted plane wave with wavevector $\#k_{t2}$.
The corresponding reflectances and transmittances are presented in
the Appendix.

\begin{figure}[!ht]
\centering \psfull
 \epsfig{file=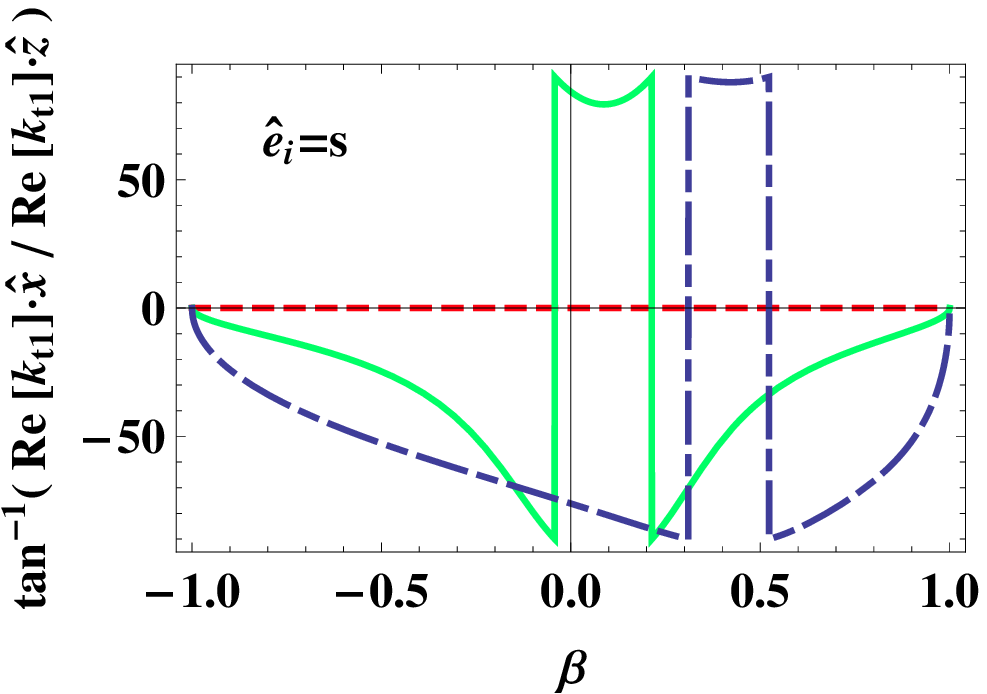,width=3.1in} \hfill
\epsfig{file=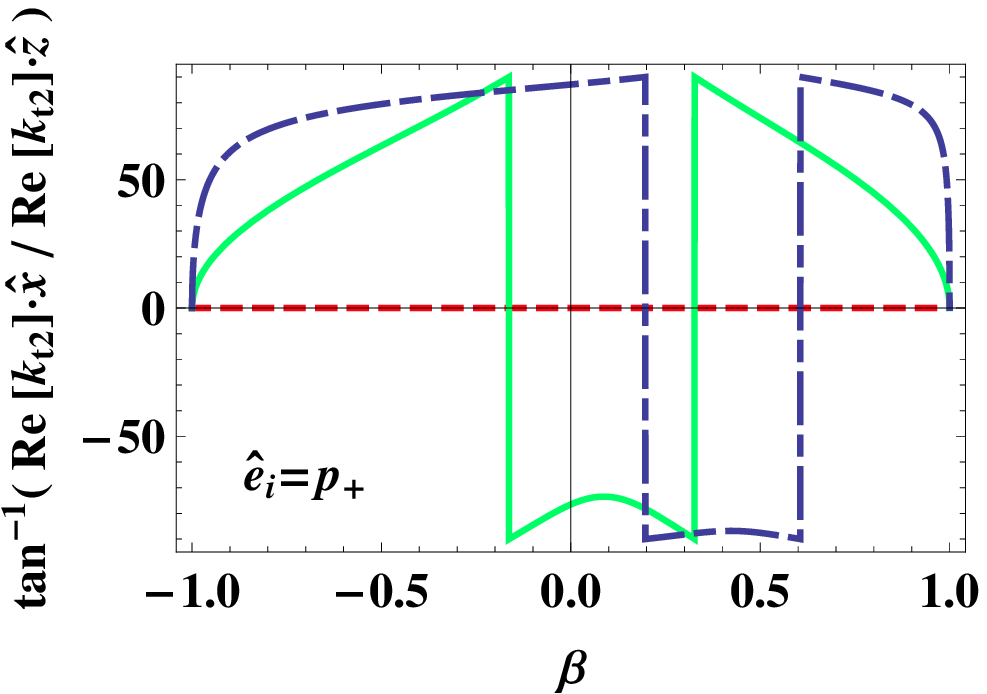,width=3.1in}
 \caption{\label{k_angle} The orientation angle of
 $\mbox{Re} \, \#k_{tj}$ (in degree)  plotted against relative speed $\beta \in \le -1, 1
 \ri$
 for the angles of incidence  $\theta = 0^\circ$   (broken curve,  red), $\theta = 5^\circ$
(solid curve,  green), and $\theta = 25^\circ $ (broken dashed
curve, blue). Plots are shown for the $s$--polarization
state ($j=1$) and  the $p$--polarization state ($j=2$) of the incident plane wave. }
\end{figure}

 The orientation angle of the real part of the wavevector $\#k_{tj}$, ($j=1,2$),
  as defined by $\tan^{-1} \le \kappa / \mbox{Re} \, k_{zj} \ri$,
 is plotted in Fig.~\ref{k_angle} versus the  relative speed $\beta \in \le -1, 1 \ri$
  for the  angle of incidence
$\theta \in \lec 0^\circ, \, 5^\circ, \, 25^\circ \ric$. For normal
incidence, we see that $\mbox{Re} \, \#k_{tj}$ is also normal to the
$z=0$ interface, for both incident $s$-- and $p$--polarization
states. However, for $\theta = 5^\circ$ and $\theta = 25 ^\circ$,
the orientation of $\mbox{Re} \, \#k_{tj}$ relative to the interface
normal is highly sensitive to $\beta$,  for both incident $s$-- and
$p$--polarization states. For example, for $\theta = 5^\circ$ with
$\hat{\#e}_i = \#s$, negative refraction occurs for $-1 < \beta <
-0.03$ and for $0.21 < \beta < 1$, and the refraction is positive
otherwise. Similarly, for $\theta = 25^\circ$ with $\hat{\#e}_i =
\#s$, and for $\theta \in\left\{5^\circ, 25^\circ\right\}$ with $\hat{\#e}_i =
\#p_+$, whether  the refraction is negative or positive depends
upon the relative speed $\beta$. The real part of $\#k_{tj}$ is
oriented normally to the $z=0$ interface in the limits $\beta \to
\pm 1$, for both $j=1$ and $2$.

\begin{figure}[!ht]
\centering \psfull
\epsfig{file=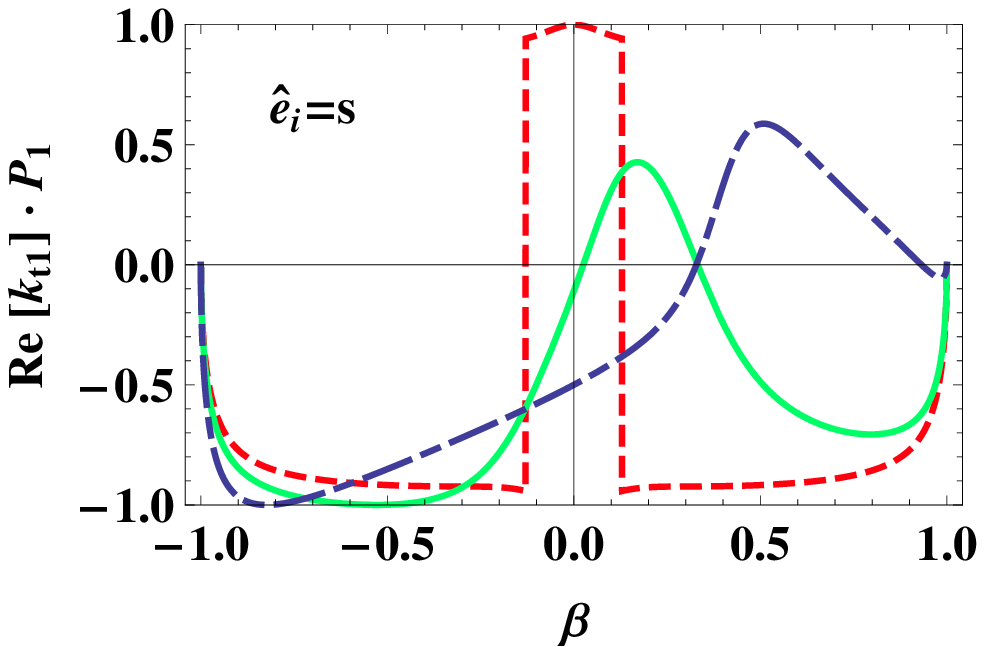,width=3.1in} \hfill
 \epsfig{file=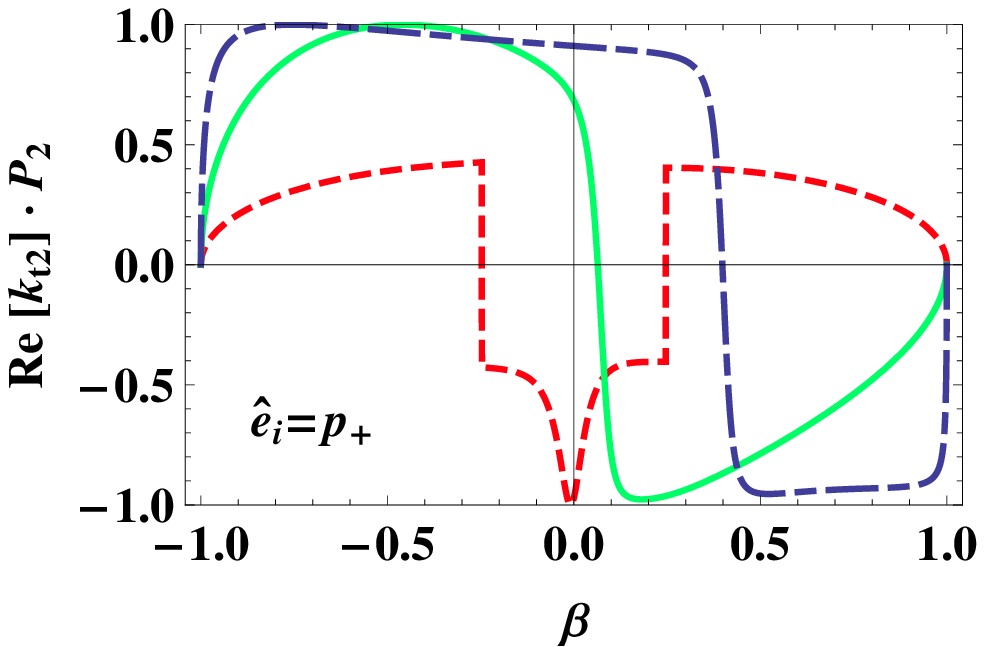,width=3.1in}
 \caption{\label{kp_fig}
As Fig.~\ref{k_angle} except that the quantity plotted against
$\beta$ is  $\#P_{j} \. \mbox{Re} \, \#k_{tj}$ (normalized),
($j=1,2$).}
\end{figure}

In Fig.~\ref{kp_fig}, the quantity $  \left. \#P_{j} \, \right|_{|
\#r | \to 0} \. \mbox{Re} \, \#k_{tj}$, ($j=1,2$), which determines
whether the phase velocity is positive or negative, is plotted
against relative speed $\beta \in \le -1, 1 \ri$
  for the  angle of incidence
$\theta \in \lec 0^\circ, \, 5^\circ, \, 25^\circ \ric$. The
numerical results echo those of Fig.~\ref{k_angle} insofar as the
sign of the phase velocity is highly sensitive to $\beta$ for both
states of incident polarization, for $\theta \in\left\{ 5^\circ, 25^\circ\right\}$.
Unlike the case of negative/positive refraction, the sign of the
phase velocity  is highly sensitive to $\beta$ even at normal
incidence, for both the $s$- and the $p$-polarization states of the incident plane wave.
 In the limits  $\beta \to \pm 1$,  $ \left. \#P_{j} \,
\right|_{| \#r | \to 0}$
 becomes oriented parallel to the interface $z=0$ for both
the $s$- and the $p$-polarization states of the incident plane wave and
 the phase velocity of the refracted plane wave
therefore becomes orthogonal.

\begin{figure}[!ht]
\centering \psfull
 \epsfig{file=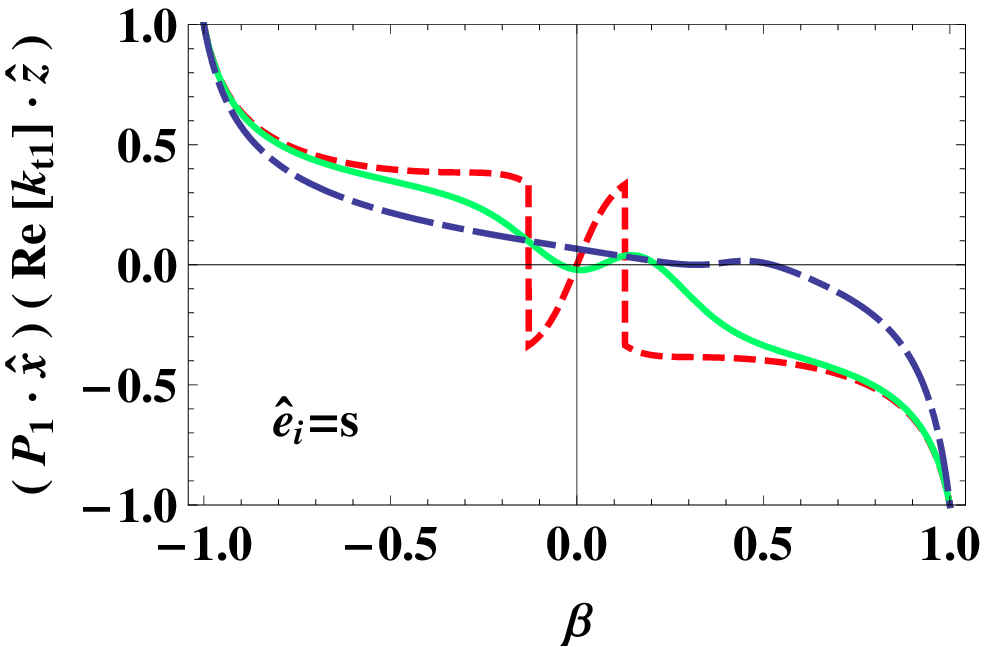,width=3.1in} \hfill
 \epsfig{file=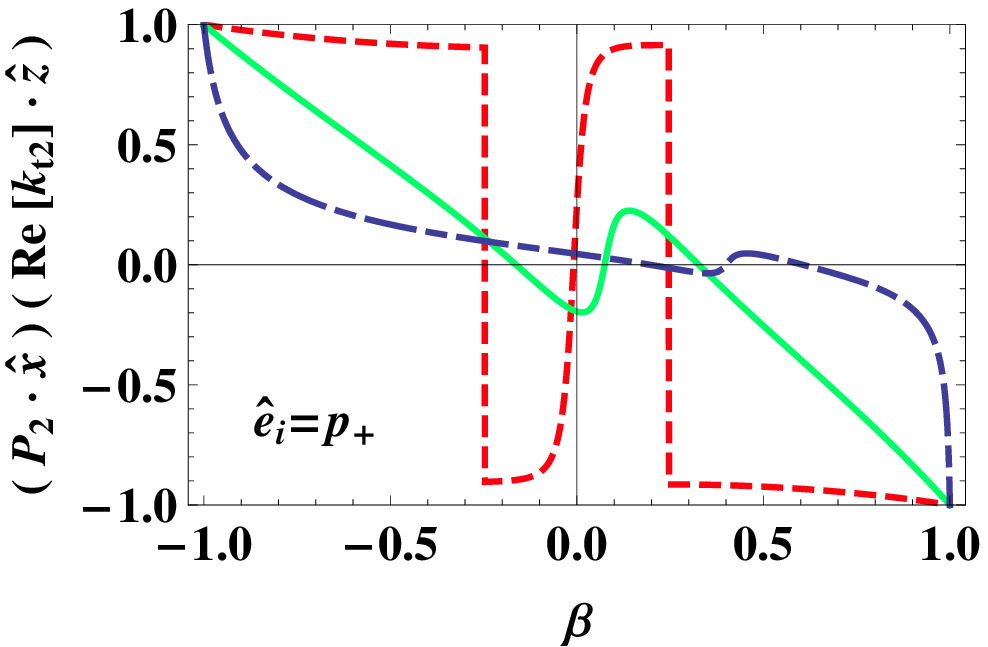,width=3.1in}
 \caption{\label{kzpx_fig}
As Fig.~\ref{k_angle} except that the quantity plotted against
$\beta$ is
 $\le \#P_{j} \. \hat{\#x} \ri  \; \mbox{Re} \, \le \#k_{tj} \. \hat{\#z}
 \ri$ (normalized), ($j=1,2$).}
\end{figure}

Lastly, we turn to the quantity $\le \#P_{j} \. \hat{\#x} \ri  \;
\mbox{Re} \, \le \#k_{tj} \. \hat{\#z} \ri$, which determines
whether or not the cycle--averaged Poynting vector and the real part
of the wavevector for the refracted plane wave are counterposed, per
the conditions \r{counter_cond}. This quantity is plotted against
relative speed $\beta \in \le -1, 1 \ri$
  for the  angle of incidence
$\theta \in \lec 0^\circ, \, 5^\circ, \, 25^\circ \ric$ in
Fig.~\ref{kzpx_fig}. In a manner similar to that represented by
Figs.~\ref{k_angle} and \ref{kp_fig}, we see that the sign of $\le
\#P_{j} \. \hat{\#x} \ri \; \mbox{Re} \, \le \#k_{tj} \. \hat{\#z}
\ri$ is highly sensitive to $\beta$ for both $j=1$ and $2$.
Furthermore, for all angles of incidence considered  with
$\hat{\#e}_i = \#s$ and  with $\hat{\#e}_i = \#p_+$, counterposition
arises in the limit $\beta \to 1$ whereas it does not arise in the
limit $\beta \to -1$.

\section{Concluding remarks}

The main conclusions to be drawn from this study are that, for the
uniformly moving pseudochiral omega material under consideration,
\begin{itemize}
\item[(i)] negative refraction is not Lorentz covariant;
\item[(ii)] negative phase velocity is not Lorentz covariant; and
\item[(iii)] counterposition is not Lorentz covariant.
\end{itemize}
While conclusions (ii) and (iii) are consistent with findings
previously reported for a uniformly moving isotropic dielectric
material \c{ML_JPA_sub,Kong_PRB}, conclusion (i) reveals a
previously unknown facet  of refraction. This has far reaching
consequences for researchers exploring the refraction of
electromagnetic waves in astrophysical applications, as well as
those investigating fundamental aspects of electromagnetic theory.
In particular, the absence of Lorentz covariance of negative
refraction further vindicates the 3+1 approach to electromagnetic
wave propagation in strong gravitational fields, which has recently
been usefully exploited to elucidate gravitationally induced
negative phase velocity in vacuum
\c{ML_JPA_GTR,MLS_NJP2,Sharif1,Komissarov}.

\section*{Appendix}
Here we present the reflectances and transmittances for the
reflection--transmission scenario described in Sec.~\ref{moving_hs}.
In terms of linearly polarized states, let the amplitude of the
electric phasor of the incident plane wave be expressed as
\begin{equation}
\#e_i = a_s \#s + a_p \#p_+,
\end{equation}
where the unit vectors $\#s = \hat{\#y}$ and $\#p_+ = - \hat{\#x}
\cos \theta + \hat{\#z} \sin \theta$. The corresponding vector
amplitude for the reflected   plane wave  may be written as
\begin{equation}
\#e_r = r_s \#s + r_p \#p_-,
\end{equation}
where the unit vector  $\#p_- =  \hat{\#x} \cos \theta + \hat{\#z}
\sin \theta$, while for the refracted plane wave we have
\begin{equation}
\#e_t = t_1 \#s + t_2 \#p,
\end{equation}
where the vector $\#p$ lies in the  plane of incidence and satisfies
$\#p \. \#p^* = 1$. The relative amplitudes $r_{s,p}/ a_{s,p}$ and
$t_{1,2}/a_{s,p}$ may be deduced by following the strategy described
in Sec.~\ref{pw_analysis}; i.e.,
 by transforming to the frame
$\Sigma'$, then invoking the boundary conditions \r{bcs}, and
finally transforming back to the frame $\Sigma$. By so doing, we
find that
\begin{equation}
\left. \begin{array}{l} a_s = 0 \implies r_s = t_1 =0 \vspace{6pt}
\\
a_p = 0 \implies r_p = t_2 =0
\end{array}
\right\}. \end{equation}
Thus, there are no cross--polarization
terms.

\begin{figure}[!ht]
\centering \psfull \epsfig{file=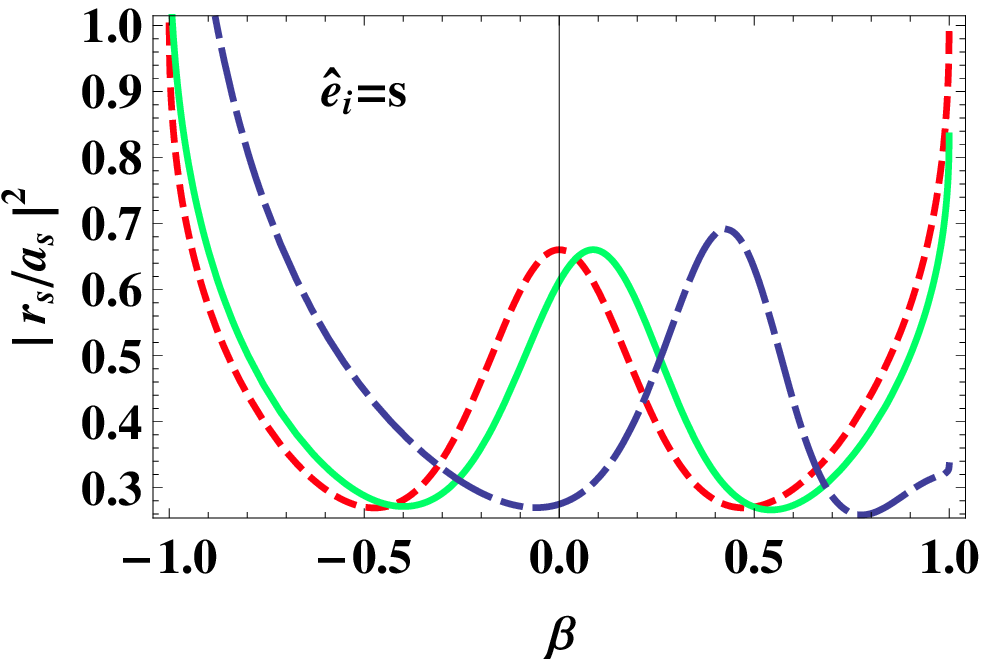,width=3.1in}
\hfill
\epsfig{file=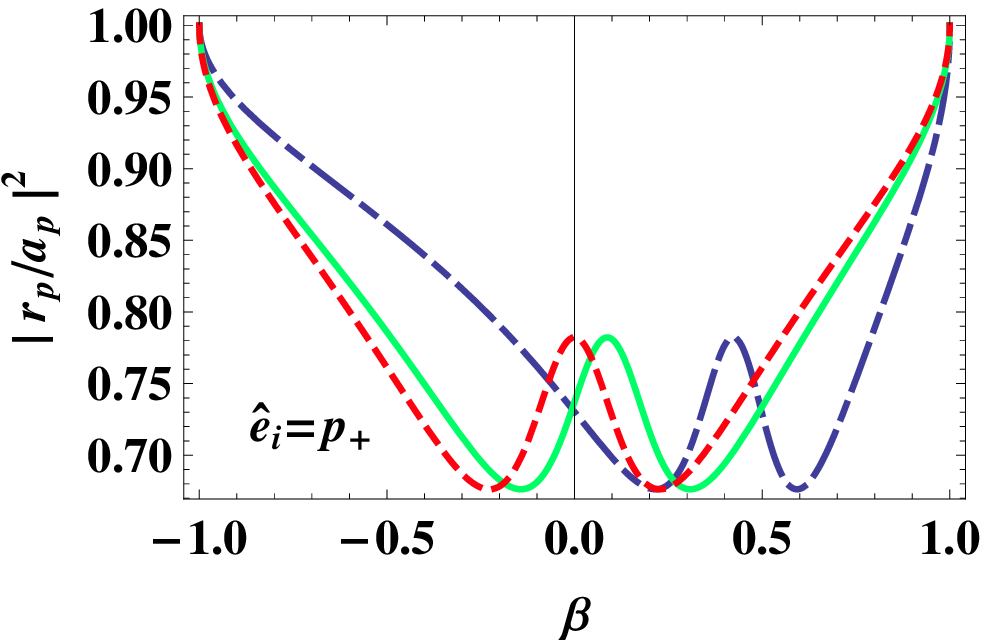,width=3.1in} \\
\epsfig{file=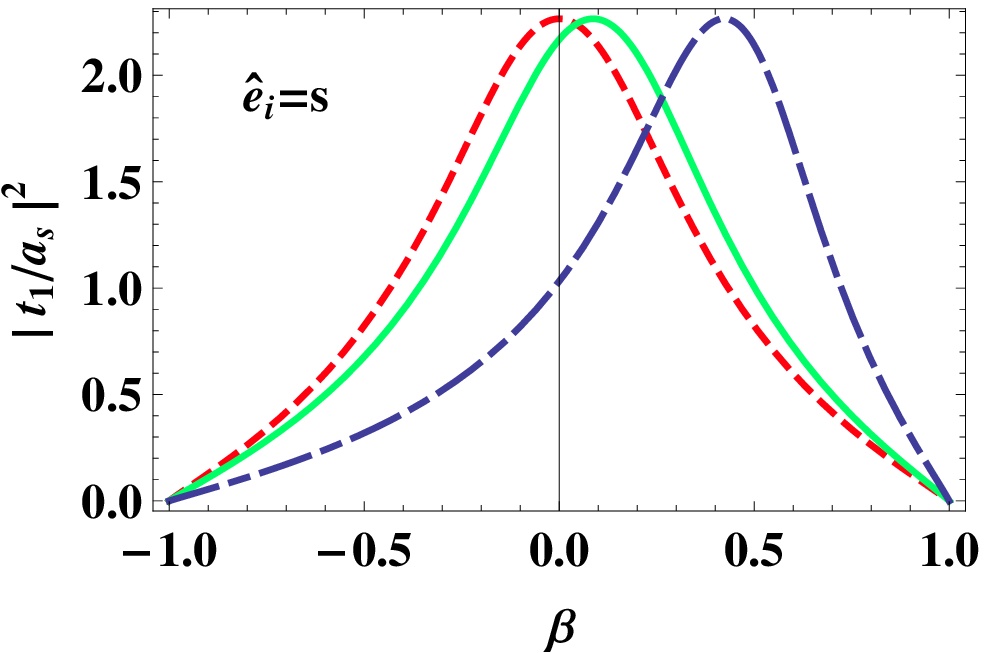,width=3.1in}  \hfill
 \epsfig{file=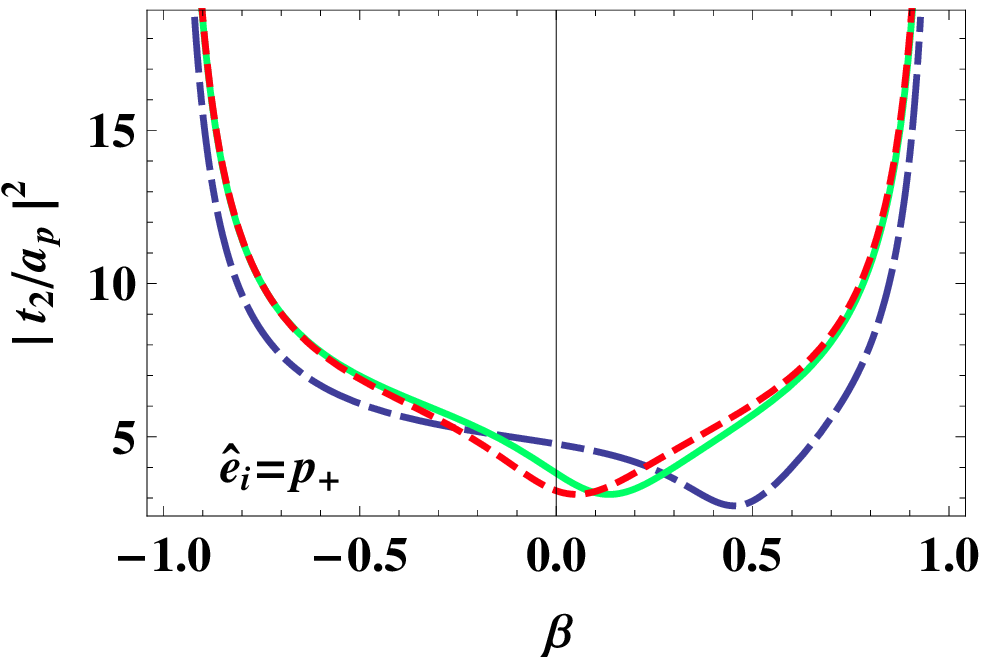,width=3.1in}
 \caption{\label{ref_trans}
As Fig.~\ref{k_angle} except that the quantities plotted against
$\beta$ are the reflectances $\left| r_{s} / a_{s} \right|^2$ and
$\left| r_{p} / a_{p} \right|^2$, and the transmittances $\left|
t_{1} / a_{s} \right|^2$ and $\left| t_{2} / a_{p} \right|^2$.
 }
\end{figure}

For the particular example considered in Sec.~\ref{num_ex},  the
reflectances $| r_{s}/ a_{p} |^2$ and $| r_{p}/ a_{p} |^2$, and the
transmittances $| t_{1}/ a_{s} |^2$ and $| t_{2}/ a_{p} |^2$, are
plotted against relative speed $\beta \in \le -1, 1 \ri$ in
Fig.~\ref{ref_trans} for the angle of incidence $\theta \in \lec
0^\circ, 5^\circ, 25^\circ \ric$. We note that, for all angles of
incidence considered,  the reflectance $| r_{p}/ a_{p} |^2$ tends to
unity in the limits $\beta \to \pm 1$ whereas this is not the case
for the reflectance $| r_{s}/ a_{s} |^2$. In a similar vein, for all
angles of incidence considered, the transmittance $| t_{1}/ a_{s}
|^2$ tends to zero in the limits $\beta \to \pm 1$ whereas the
transmittance $| t_{2}/ a_{p} |^2$ does not.

\newpage

\vspace{15mm}

%\vspace{15mm}
\newpage

\vspace{15mm}

\begin{thebibliography}{99}
\bibitem{note1}{The phenomenon of counterposition is called  `negative
refraction' in Ref.~\cite{Kong_PRB}. In contrast, we adopt the
standard convention wherein the relative orientations of the real
parts of the
 refraction and incidence wavevectors alone
determine the sense of refraction,  in consonance with Snel's law
\c{Chen}.}
\end{thebibliography}
\end{document}